\documentclass[final,number,sort&compress,times,3p,12pt]{elsarticle}

\usepackage{graphicx}
\usepackage{dcolumn}
\usepackage{bm}
\usepackage{epstopdf}
\usepackage{bm,hyperref,color,breakurl}

\journal{Journal of Magnetism and Magnetic Materials}

	\usepackage{fancyheadings}
\renewcommand{\thispagestyle}[1]{} 

\begin{document}


\title{Self-consistent model of a solid \\for the description of lattice and magnetic properties}


\author[a1]{T. Balcerzak}
\ead{t\_balcerzak@uni.lodz.pl}
\author[a1]{K. Sza\l{}owski\corref{cor1}}
\ead{kszalowski@uni.lodz.pl}

\author[a3]{M. Ja\v{s}\v{c}ur}
\address[a1]{Department of Solid State Physics, Faculty of Physics and Applied Informatics,\\
University of \L\'{o}d\'{z}, ulica Pomorska 149/153, 90-236 \L\'{o}d\'{z}, Poland}

\address[a3]{Department of Theoretical Physics and Astrophysics, Faculty of Science, \\P. J. \v{S}\'{a}f\'{a}rik University, Park Angelinum 9, 041 54 Ko\v{s}ice, Slovak Republic}

\cortext[cor1]{Corresponding author}


\date{\today}

\begin{abstract}
In the paper a self-consistent theoretical description of the lattice and magnetic properties of a model system with magnetoelastic interaction is presented. The dependence of magnetic exchange integrals on the distance between interacting spins is assumed, which couples the magnetic and the lattice subsystem. The framework is based on summation of the Gibbs free energies for the lattice subsystem and magnetic subsystem. On the basis of minimization principle for the Gibbs energy, a set of equations of state for the system is derived. These equations of state combine the parameters describing the elastic properties (relative volume deformation) and the magnetic properties (magnetization changes). 

The formalism is extensively illustrated with the numerical calculations performed for a system of ferromagnetically coupled spins $S$=1/2 localized at the sites of simple cubic lattice. In particular, the significant influence of the magnetic subsystem on the elastic properties is demonstrated. It manifests itself in significant modification of such quantities as the relative  volume deformation, thermal expansion coefficient or isothermal compressibility, in particular, in the vicinity of the magnetic phase transition. On the other hand, the influence of lattice subsystem on the magnetic one is also evident. It takes, for example, the form of dependence of the critical (Curie) temperature and magnetization itself on the external pressure, which is thoroughly investigated. 
\end{abstract}



\maketitle

\section{Introduction}

Thermodynamics of magnetic solids is a subject of interest of solid state physicists since many years \cite{Stanley}. From the point of view of methodology, some analogy to systems described by the volume and pressure is exploited, namely the thermodynamic magnetic variables: magnetic field $h$ and magnetization $m$ correspond to the respective mechanical variables - pressure $p$ and volume $V$, considered at some temperature $T$ (see for example Ref.~\cite{Balcerzak1998}). Therefore, the magnetic equation of state involves three variables: $h$, $m$ and $T$. 

In majority of cases both the subsystems of solid state (magnetic and lattice one) are described separately, with no coupling between them. In such situation, the lattice-related properties of the magnetic solid are considered as fully independent on magnetic properties, leading to another equation of state interrelating $p$, $V$ and $T$. However, this approach neglects the magnetoelastic interactions which occur between these subsystems. The simplest source of this kind of coupling is the fact that the magnetic exchange integral between magnetic moments depends on their mutual distance, thus is a volume-dependent quantity. The magnetoelastic interactions are basis for such effects as the magnetostriction and piezomagnetism, which are important from the point of view of possible applications. They are also responsible for sensitivity of any magnetic properties (for example, the critical temperature) to external pressure.

Among the literature concerning the studies of magneto-elastic interactions, many particular contributions can be mentioned. One of the subjects of intensive studies was compressible Ising model of various dimensionalities  \cite{Salinas1973,Bergman1976,Chakrabarti1977,Chakrabarti1979,Henriques1987,Diep1999,Diep2000,Boubcheur2001,Landau2005,Li2010}, including the diluted case \cite{Chakrabarti1980,Chakrabarti1982} and the quantum versions \cite{Chakrabarti1980b}. The studies involved also Heisenberg model \cite{Pytte1965,Barma,Diep2002} or spin glasses \cite{Liarte} or highly interesting frustrated magnetic systems \cite{Sobkowicz1996,Vecchini,Zorko}. Some of the results also incorporate magnetoelastic coupling into exactly solvable models \cite{Bergman1973,Strecka2012,Strecka2012b}. In another approach, the pressure influence on the Curie temperature has been discussed \cite{Wojtczak,Leger}. Also, an attempt has been undertaken aimed at describing the quantum phase transitions, triggered by the external pressure \cite{Gehring}. Among studies of specific materials, the example of EuTe can be mentioned, for which the influence of pressure on magnetic phase diagram has attracted both theoretical (\cite{Radomska2000,Radomska2001,Radomska2003}) and experimental \cite{Sollinger2010} attention. It should be mentioned that one of the approaches to theoretical description of systems with magnetoelastic coupling is based on the Landau theory of phase transitions, which involves a semi-empirical expression for free energy \cite{Amaral2004,Sigrist2005}. Some approaches involve the additional magnetoelastic terms in expressions for molecular field \cite{Alho2010}. Another method, including structural transformations, was used in Monte Carlo studies of Refs.~\cite{Singh2013}. These approaches proved their usefulness in characterization of the magnetocaloric effect.

However, in spite of the existence and usefulness of various developed formalisms, there is still a room for fully microscopic approach, based on a full Hamiltonian and allowing to construct complete thermodynamic description of the system in question. Such a general theory, based on the statistical thermodynamic approach and capable of describing both interacting subsystems (lattice and magnetic one) in a complete, self-consistent way would be of a great value. 

Motivated by this situation, the present paper is aimed at developing the method which will be suitable for the complete, fully consistent, statistical-thermodynamic description of the model solid state, with magnetoelastic interaction taken into account. 

The key point is to obtain the Gibbs free-energy in the most general form, from which all the interesting thermodynamic quantities (both magnetic and non-magnetic ones), as well as the relationships between them, can be obtained. Employing the variational principle for the Gibbs energy, with respect to the volume and magnetization, the set of two coupled equations of state will be derived.

The method will enable calculations of the magnetic quantities, such as magnetization and Curie temperature in the presence of external pressure, as well as the non-magnetic (structural) quantities, such as: volume, compressibility or thermal expansion in the magnetic field.\\

The paper is organized as follows: The theoretical approach will be presented in the next Section \ref{sec2}. Some complementary mathematical formalism concerning this Section will be included in the Appendices: \ref{appendix1} and \ref{appendix2}.  In the Section \ref{sec3} the results of numerical calculations will be presented in the figures and discussed. Finally, in the last Section \ref{sec4} the summary will be presented and some conclusions will be drawn.\\

\section{Theoretical model}
\label{sec2}
The Gibbs free energy of a system is assumed in the form of:\\
\begin{equation}
\label{eq1}
G=G_V+G_m
\end{equation}
where $G_V$ and $G_m$ are the Gibbs energies of non-magnetic (lattice) and magnetic subsystems, respectively. \\

\subsection{Lattice subsystem}

The Gibbs energy for non-magnetic subsystem is composed of the following parts:
\begin{equation}
\label{eq2}
G_V=F_{\varepsilon}+F_{\rm D}+pV,
\end{equation}
where $F_{\varepsilon}$ is the elastic (static) energy, $F_{\rm D}$ is the vibrational (thermal) energy in Debye approximation and $p$ is the external pressure.\\
The elastic energy can be found basing on the Morse potential \cite{Morse,Girifalco,Lincoln}:
\begin{equation}
\label{eq3}
U(r)=D\left(1-e^{-\alpha \left(r-r_0\right)/r_0}\right)^2
\end{equation}
which contains three fitting parameters: potential depth $D$, dimensionless asymmetry parameter $\alpha$ and the distance $r_0$ where the potential has its minimum.\\
For the crystals with cubic symmetry the interatomic distance $r$ can be expressed in terms of the isotropic volume deformation $\varepsilon$, namely:
\begin{equation}
\label{eq4}
r=r_{j,0}\left(1+\varepsilon\right)^{1/3}
\end{equation}
where $\varepsilon$ is defined by the equation:
\begin{equation}
\label{eq5}
V=V_0\left(1+\varepsilon\right)
\end{equation}
and $V_0 = V(p=0, T=0)$ is the volume of non-deformed system at $p=0$ and $T=0$. $r_{j,0}$ in Eq.(\ref{eq4}) is the interatomic distance between the central and $j$-th atom in non-deformed crystal.\\
It is convenient to shift the elastic potential by a constant value in order to set zero energy $F_{\varepsilon}(\varepsilon=0)=0$ for non-deformed crystal. Then, for $N$ atoms in a sample,  the elastic energy can be written as a sum over all interacting pairs:
\begin{equation}
\label{eq6a}
F_{\varepsilon}=\frac{N}{2}D\sum_{j}^{}{\{ \left[1-e^{-\alpha \left(\frac{r_{j,0}}{r_0}\left(1+\varepsilon \right)^{1/3}-1\right)}\right]^2 - 
\left[1-e^{-\alpha \left(\frac{r_{j,0}}{r_0}-1\right)}\right]^2 \}}
\end{equation}
The summation accounts for the long-range interactions in the Morse potential. The sum in Eq.(\ref{eq6a}) can be performed over the coordination zones with radius $r_{k,0}$ and  coordination numbers $z_k$. Thus, we present Eq.(\ref{eq6a}) in the form of:
\begin{equation}
\label{eq6}
F_{\varepsilon}=\frac{N}{2}D\sum_{k}^{}{z_k\{ \left[1-e^{-\alpha \left(\frac{r_{1,0}}{r_0}\frac{r_{k,0}}{r_{1,0}}\left(1+\varepsilon \right)^{1/3}-1\right)}\right]^2 - 
\left[1-e^{-\alpha \left(\frac{r_{1,0}}{r_0}\frac{r_{k,0}}{r_{1,0}}-1\right)}\right]^2 \}}
\end{equation}
where  $r_{k,0}/r_{1,0}$ and $z_k$ can be found numerically for given crystallographic structure.
The equilibrium nearest-neighbour (NN) normalized distance $\frac{r_{1,0}}{r_0}$ will be determined later from the minimum of the total energy.
The expression (\ref{eq6}) is then convenient for use for arbitrary isotropic deformation $\varepsilon$.\\
The elastic energy is a source of static pressure:
\begin{eqnarray}
\label{eq7}
p_{\varepsilon}&=&-\left(\frac{\partial F_{\varepsilon}}{\partial V}\right)_T=-\frac{1}{V_0}\left(\frac{\partial F_{\varepsilon}}{\partial \varepsilon}\right)_T=\nonumber\\
&=&-\frac{1}{3}\frac{N}{V_0}D\alpha \frac{r_{1,0}}{r_0}\sum_{k=1}^{}{z_k \frac{r_{k,0}}{r_{1,0}}\left[1-e^{-\alpha \left(\frac{r_{1,0}}{r_0}\frac{r_{k,0}}{r_{1,0}}\left(1+\epsilon \right)^{1/3}-1\right)}\right]
\frac{e^{-\alpha \left(\frac{r_{1,0}}{r_0}\frac{r_{k,0}}{r_{1,0}}\left(1+\varepsilon \right)^{1/3}-1\right)}}{\left(1+\varepsilon \right)^{2/3}}},
\end{eqnarray}
which, together with other pressure contributions, keeps the system in equilibrium.\\

The vibrational energy is taken in the Debye approximation and its form can appear in two variants: for low temperatures only, and in the whole temperature range.\\
In the low temperature limit the free energy is given by the formula \cite{Balcerzak}:
\begin{equation}
\label{eq8}
F_{\rm D}=N\left[\frac{9}{8}k_{\rm B}T_{\rm D}-\frac{1}{5}\pi^4 k_{\rm B}T\left(\frac{T}{T_{\rm D}}\right)^3\right].
\end{equation}
The Debye temperature $T_{\rm D}$ is volume-dependent and can be presented in the approximate form \cite{Matsui}:
\begin{equation}
\label{eq9}
T_{\rm D}=T_{\rm D}^0 e^{\left(\gamma_{\rm D}^0-\gamma_{\rm D}\right)/q}=T_{\rm D}^0
e^{\gamma_{\rm D}^0\left[1-\left(1+\varepsilon \right)^q\right]/q}
\end{equation}
where the  Gr\"uneisen parameter $\gamma_{\rm D}$ is given by \cite{Gruneisen}:
\begin{equation}
\label{eq10}
\gamma_{\rm D}=-\frac{V}{T_{\rm D}}\left(\frac{\partial T_{\rm D}}{\partial V}\right)_T= \gamma_{\rm D}^0\left(1+\varepsilon \right)^q.
\end{equation}
$T_{\rm D}^0$ and $\gamma_{\rm D}^0$ are the Debye temperature and Gr\"uneisen parameter, respectively, which are taken at $T=0$ and $p=0$. It has been shown that for the Morse potential the Gr\"uneisen parameter $\gamma_{\rm D}^0$ can be expressed as
\cite{Krivtsov2011}:
\begin{equation}
\label{eq10a}
 \gamma_{\rm D}^0=\left(3\alpha-2\right)/6,
 \end{equation}
which, via elastic potential parameters, introduces anharmonicity to the Debye model.\\
The Debye energy (Eq.(\ref{eq8})) gives rise to the vibrational pressure for low temperatures:
\begin{equation}
\label{eq11}
p_{\rm D}=-\left(\frac{\partial F_{\rm D}}{\partial V}\right)_T=3\frac{N}{V_0}k_{\rm B}T_{\rm D}\gamma_{\rm D}\left[\frac{3}{8}+\frac{1}{5}\pi^4 \left(\frac{T}{T_{\rm D}}\right)^4 \right]\frac{1}{1+\varepsilon}
\end{equation}
where $T_{\rm D}$ is given by Eq.(\ref{eq9}), and $\left(\partial T_{\rm D}/\partial V\right)_T$ is expressed on the basis of Eq.(\ref{eq10}).\\
In general, for any temperature, the vibration energy can be found from the formula \cite{Wallace}:
\begin{equation}
\label{eq12}
F_{\rm D}=N\left[\frac{9}{8}k_{\rm B}T_{\rm D}+3k_{\rm B}T\ln \left(1-e^{-y_{\rm D}}\right)-3k_{\rm B}T\frac{1}{y_{\rm D}^3}\,\int_{0}^{y_{\rm D}}{\frac{y^3}{e^{y}-1}dy}\right].
\end{equation}
where $y_{\rm D}=T_{\rm D}/T$.  

Such energy gives the following vibrational pressure:
\begin{equation}
\label{eq13}
p_{\rm D}=-\left(\frac{\partial F_{\rm D}}{\partial V}\right)_T=9\frac{N}{V_0}k_{\rm B}T_{\rm D}\gamma_{\rm D}       
\left[\frac{1}{8}+\frac{1}{y_{\rm D}^4} \int_{0}^{y_{\rm D}}{\frac{y^3}{e^{y}-1}\,dy}\right] \frac{1}{1+\varepsilon}
\end{equation}
It can be noted that the integral in Eqs.(\ref{eq12}) and (\ref{eq13}) can be calculated for $T>T_{\rm D}/(2\pi)$ by the approximate method (as in Ref.\cite{Balcerzak}) using Bernoulli series.
On the other hand, for the whole temperature range the integral can be calculated either by the direct numerical  integration, or by the exact method as, for instance, presented in Ref.\cite{Dubinov2008}, using special functions. In the exact method one can use the following formula \cite{Wood} (see \ref{appendix1}):
\begin{eqnarray}
\label{eq13a}
\int_{0}^{y_{\rm D}}{\frac{y^3}{e^{y}-1}\,dy}&=&
\frac{1}{15}\pi^4-3!\sum_{k=0}^{3}{{\rm Li}_{4-k}\left(e^{- y_{\rm D}}\right)
\frac{y_{\rm D}^k}{k!}}\nonumber\\&=&
\frac{1}{15}\pi^4+y_{\rm D}^3 \ln \left(1- e^{-y_{\rm D}}\right)\nonumber\\
&-&3y_{\rm D}^2{\rm Li}_{2}\left(e^{-y_{\rm D}}\right) - 6y_{\rm D}{\rm Li}_{3}\left(e^{-y_{\rm D}}\right)-6{\rm Li}_{4}\left(e^{-y_{\rm D}}\right),
\end{eqnarray}
where ${\rm Li}_{s}\left(z\right)=\sum_{k=1}^{\infty}z^k/k^s$ is the polylogarithm of order $s$ and argument $z$, extended by the process of analytic continuation. Substitution of the above formula into Eqs.(\ref{eq12}) and (\ref{eq13}) leads to the expressions:
\begin{eqnarray}
\label{eq13b}
F_{\rm D}&=&N\left\{ \frac{9}{8}k_{\rm B}T_{\rm D}-\frac{1}{5}\pi^4 k_{\rm B}T\frac{1}{y_{\rm D}^3} \nonumber   \right. \\ &&\left. + 9k_{\rm B}T\frac{1}{y_{\rm D}}\left[
{\rm Li}_{2}\left(e^{-y_{\rm D}}\right) +\frac{2}{y_{\rm D}}{\rm Li}_{3}\left(e^{-y_{\rm D}}\right)+\frac{2}{y_{\rm D}^2}{\rm Li}_{4}\left(e^{-y_{\rm D}}\right)
\right] \right\},
\end{eqnarray}
and
\begin{eqnarray}
\label{eq13c}
p_{\rm D}&=&3\frac{N}{V_0}k_{\rm B}T_{\rm D}\gamma_{\rm D}       
\left\{ \frac{3}{8}+\frac{1}{5}\pi^4 \frac{1}{y_{\rm D}^4} +\frac{3}{y_{\rm D}}\ln \left(1- e^{-y_{\rm D}}\right) \right. \nonumber \\ &&\left.
-\frac{9}{y_{\rm D}^2}\left[{\rm Li}_{2}\left(e^{-y_{\rm D}}\right) +\frac{2}{y_{\rm D}}{\rm Li}_{3}\left(e^{-y_{\rm D}}\right)+\frac{2}{y_{\rm D}^2}{\rm Li}_{4}\left(e^{-y_{\rm D}}\right)
\right] \right\} \frac{1}{1+\varepsilon}.
\end{eqnarray}
It is interesting to note that the above equations (\ref{eq13b}) and (\ref{eq13c}) present a generalization of the corresponding equations  (\ref{eq8}) and (\ref{eq11}), known from conventional  low-temperature approximation for the Debye model. The last formulas  are valid for arbitrary temperature, including $T \to 0$ limit, which can be proved on the basis of the relation:
${\rm lim}_{|z|\to 0}{\rm Li}_s\left(z\right)=z$. \\

\subsection{Magnetic subsystem}

As far as the magnetic free energy is concerned, its simplest form follows from the Molecular Field Approximation (MFA) which we apply here for arbitrary spin $S$ and the long-range exchange interactions. The magnetic Gibbs energy is then given by \cite{Szalowski}:
\begin{equation}
\label{eq14}
 G_m=-Nk_{\rm B}T \ln \{ \frac{\sinh \left[\frac{2S+1}{2}\beta\left(m\displaystyle\sum_{k}^{}{ J_k z_k+h}\right) \right]}{\sinh \left[\frac{1}{2}\beta\left(m\displaystyle\sum_{k}^{}{ J_k z_k+h}\right) \right]} \} + \frac{N}{2}m^2 \sum_{k}^{}{ J_k z_k},
\end{equation}
where $m$ is on-site magnetization, $h$ stands for the external magnetic field and $z_k$ is the number of spins on the $k$-th coordination zone. The exchange integral $J_k=J\left(r_k\right)$ is the exchange integral for the $k$-th zone of radius $r_k$. In the present formulation we assume the ferromagnetic coupling, i.e. $J\left(r\right)>0$ to deal with a magnetic system which does not need to be subdivided into magnetic sublattices, so that the values of magnetization $m$ are equal at every lattice site.

We can relate the distance dependence of the exchange integral to the volume dependence via formulas based on Eq.(\ref{eq4}), namely:
\begin{eqnarray}
\label{eq15}
r_{1,\rm C}=r_{1,0}\left(1+\varepsilon_{\rm C}\right)^{1/3}\nonumber\\
r_{k}=r_{k,0}\left(1+\varepsilon \right)^{1/3}
\end{eqnarray}
where $r_{k,0}$ is the radius of $k$th coordination zone in non-perturbed system, when $p=0$, $h=0$, and $T=0$. This notation is in agreement with Eqs.(\ref{eq6}) and (\ref{eq7}), and $\varepsilon_{\rm C}$ corresponds here to the volume deformation at $p=0$, $h=0$, and critical (Curie) temperature $T=T_{\rm C}$. The constant deformation parameter $\varepsilon_{\rm C}$ will be determined later.\\
It should be strongly emphasized that the values of exchange integrals $J_k$ are lattice deformation-dependent, what couples the magnetic and lattice subsystems.

The first equation of state can be derived from the minimum condition for the total Gibbs energy (\ref{eq1}) with respect to $m$ treated as a variational parameter:
\begin{equation}
\label{eq23}
\frac{\partial G}{\partial m}=0.
\end{equation}
This condition yields the relationship:
\begin{equation}
\label{eq22}
m=SB_S\left[S\beta \left(m\sum_{k} J_k z_k+h\right) \right],
\end{equation}
where $SB_S\left(Sx\right)$ is the Brillouin function:
\begin{equation}
\label{eq22a}
SB_S\left(Sx\right)=\frac{2S+1}{2}\coth\left(\frac{2S+1}{2}x\right)-
\frac{1}{2}\coth\left(\frac{x}{2}\right).
\end{equation}
From the free energy (\ref{eq14}), the magnetic contribution to the pressure can be found:
\begin{equation}
\label{eq21}
p_{m}=-\left(\frac{\partial G_m}{\partial V}\right)_T=\frac{1}{2}\frac{N}{V_0}m^2
 \sum_{k} \frac{\partial J_{k}}{\partial \epsilon} z_k.
\end{equation}
The derivative of the exchange integral with respect to the relative deformation yields:
\begin{equation}
\label{eq21a}
\frac{\partial J_k}{\partial \epsilon} = \frac{\partial J\left(r_k/r_{1,\rm C}\right)}{\partial \epsilon} = \frac{\partial J\left(r_k/r_{1,\rm C}\right)}{\partial\left(r_k/r_{1,\rm C}\right)} \frac{1}{r_{1,\rm C}}\frac{\partial r_k}{\partial \epsilon}  = \frac{1}{3\left(1+\epsilon\right)^{2/3}\left(1+\epsilon_{\rm C}\right)^{1/3}}\frac{r_{k,0}}{r_{1,0}}\frac{\partial J\left(r_k/r_{1,\rm C}\right)}{\partial \left(r_k/r_{1,\rm C}\right)}.
\end{equation}
For brevity we denote $\frac{\partial J\left(r_k/r_{1,\rm C}\right)}{\partial \left(r_k/r_{1,\rm C}\right)}=J'_{k}$ and finally we obtain:
\begin{equation}
\label{eq21b}
p_{m}=-\left(\frac{\partial G_m}{\partial V}\right)_T=\frac{1}{6}\frac{N}{V_0}m^2\frac{1}{\left(1+\epsilon\right)^{2/3}\left(1+\epsilon_{\rm C}\right)^{1/3}}
 \sum_{k}^{}{\frac{r_{k,0}}{r_{1,0}} J'_{k} z_k}.
\end{equation}
The second equation of state results from the analogous minimum condition with respect to variable $\varepsilon$:
\begin{equation}
\label{eq24}
\frac{\partial G}{\partial \epsilon}=0,
\end{equation}
which leads to the relationship:
\begin{equation}
\label{eq25}
p_{\varepsilon}+p_{\rm D}+p_{m}=p.
\end{equation}
In Eq.(\ref{eq25}) $p$ is the external pressure, and $p_{\varepsilon}$, $p_{\rm D}$, and $p_{m}$ are given by Eq.(\ref{eq7}), Eq.(\ref{eq11}) for low temperatures only or Eq.(\ref{eq13c}) for arbitrary temperature, and Eq.(\ref{eq21b}), respectively.\\
From Eq.(\ref{eq22}) the phase transition (Curie) temperature can be found, when we  put $h=0$ and $m \to 0$:
\begin{equation}
\label{eq26}
k_{\rm B}T_{\rm C}=\frac{S\left(S+1\right)}{3}\sum_{k} J_k z_k,
\end{equation}
where it should be remembered that the values of the exchange integrals $J_{k}$ should be taken at the appropriate relative deformation $\varepsilon=\varepsilon \left(m=0, p\right)$. In particular case, when $p=0$, then $\varepsilon=\varepsilon_{\rm C}$.\\
Equations of state can be first analysed for $p=0$, $h=0$, and two characteristic temperatures: $T=T_{\rm C}$ and $T=0$. For $T \to T_{\rm C}$ the magnetic pressure $p_m$ vanishes on the basis of Eq.(\ref{eq21b}), and from Eq.(\ref{eq25}) we get:
\begin{equation}
\label{eq27}
p_{\varepsilon}+p_{\rm D}=0,
\end{equation}
where $p_{\rm D}$ is given by Eq.(\ref{eq13c}), and the Curie temperature in Eq.(\ref{eq27}) is taken from the formula (\ref{eq26}) for $p=0$. Thus, Eq.(\ref{eq27}) takes the form of:
\begin{eqnarray}
\label{eq27a}
&&\frac{1}{3}\frac{D}{k_{\rm B}T_{\rm D}^0}\alpha \frac{r_{1,0}}{r_0}\sum_{k}^{}{z_k \frac{r_{k,0}}{r_{1,0}}\left[1-e^{-\alpha \left(\frac{r_{1,0}}{r_0}\frac{r_{k,0}}{r_{1,0}}\left(1+\varepsilon_{\rm C} \right)^{1/3}-1\right)}\right]
\frac{e^{-\alpha \left(\frac{r_{1,0}}{r_0}\frac{r_{k,0}}{r_{1,0}}\left(1+\varepsilon_{\rm C} \right)^{1/3}-1\right)}}{\left(1+\varepsilon_{\rm C} \right)^{2/3}}}=\nonumber\\
&&
3\frac{T_{\rm D}}{T_{\rm D}^0}\gamma_{\rm D}\left\{
 \frac{3}{8}+\frac{1}{5}\pi^4 \frac{1}{y_{\rm D}^4} +\frac{3}{y_{\rm D}}\ln \left(1- e^{-y_{\rm D}}\right) \right. \nonumber \\ &&\left.
-\frac{9}{y_{\rm D}^2}\left[{\rm Li}_{2}\left(e^{-y_{\rm D}}\right) +\frac{2}{y_{\rm D}}{\rm Li}_{3}\left(e^{-y_{\rm D}}\right)+\frac{2}{y_{\rm D}^2}{\rm Li}_{4}\left(e^{-y_{\rm D}}\right)
\right] 
\right\}\frac{1}{1+\varepsilon_{\rm C}}
\end{eqnarray}
whereas $T=T_{\rm C}$ for $p=0$, and $T_{\rm D}$ and $\gamma_{\rm D}$ are taken at $\varepsilon=\varepsilon_{\rm C}$.\\
In turn, for $T =0$ and $h=0$, from Eq.(\ref{eq22}) we obtain $m=S$, independently on $J_k$. Then, the magnetic pressure amounts to:
\begin{equation}
\label{eq28}
p_m = \frac{1}{6}\frac{N}{V_0}S^2\frac{1}{\left(1+\epsilon_{\rm C}\right)^{1/3}}\sum_{k} \frac{r_{k,0}}{r_{1,0}}J'_{k} z_k ,
\end{equation}
where for $p=0$ and $T=0$ we assume  $\varepsilon=0$. Lack of deformation also simplifies the expressions for $p_{\varepsilon}$ (Eq.(\ref{eq7})) and $p_{\rm D}$ (Eq.(\ref{eq11}) or (\ref{eq13c})). Thus,  the equation (\ref{eq25}) takes the following form in the ground state:
\begin{eqnarray}
\label{eq29}
&&\frac{1}{3}\frac{D}{k_{\rm B}T_{\rm D}^0}\alpha \frac{r_{1,0}}{r_{0}}
\sum_{k}^{}{z_k \frac{r_{k,0}}{r_{1,0}}\left[1-e^{-\alpha \left(\frac{r_{1,0}}{r_0}\frac{r_{k,0}}{r_{1,0}}-1\right)}\right]
e^{-\alpha \left(\frac{r_{1,0}}{r_0}\frac{r_{k,0}}{r_{1,0}}-1\right)}}\nonumber\\
&&- 
\frac{1}{6}\frac{1}{k_{\rm B}T_{\rm D}^0}S^2 \frac{1}{\left(1+\epsilon_{\rm C}\right)^{1/3}}\sum_{k} \frac{r_{k,0}}{r_{1,0}}J'_{k} z_k 
=\frac{9}{8}\gamma_{\rm D}^0.
\end{eqnarray}

From the set of those two equations of state, (\ref{eq27a}) and (\ref{eq29}), the constant deformation parameter $\varepsilon_{\rm C}$ 
and the equilibrium NN distance, i.e., $r_{1,0}/r_0$ ratio,
can simultaneously be determined. Knowledge of these two constants enables further calculations based on the general equations of state (\ref{eq22}) and (\ref{eq25}), for arbitrary temperature $T$, external pressure $p$ and magnetic field $h$.

A case of special interest is the one with magnetic interactions limited to nearest-neighbours only. Moreover, the interaction can be assumed to follow the power law as a function of the distance between nearest-neighbour spins. Let us mention that such a form of the distance dependence of exchange integral has been found experimentally for example in neutron scattering studies of magnetic semiconductors \cite{Wiren,Szuszkiewicz2006}. The specific form of the appropriate equations for that case is presented in detailed form in \ref{appendix2}.

\section{Numerical results and discussion}
\label{sec3}

In order to illustrate our formalism for general ferromagnetic system, we have selected a model solid based on the three-dimensional simple cubic (sc) lattice. Each lattice site carries localized spin $S=$1/2 and nearest-neighbour spins interact ferromagnetically, with the exchange coupling energy varying with the interspin distance according to a power law. This is exactly the case described in \ref{appendix2}. In the present section we discuss the extensive calculations of magnetic and lattice-related properties for the described model. The calculations are based on a pair of equations of state given by Eqs.~\ref{eq27a} and \ref{eq35}. It should be stressed that, prior to calculations based on equations \ref{eq25} and \ref{eq35}, the parameters $r_{1,0}/r_{0}$ and $\epsilon_{\rm C}$ have to be determined from the equations \ref{eq27a} and \ref{eq34b}. Regarding elastic interactions the summation over the coordination zones of the sc lattice in Eq.~\ref{eq25} and \ref{eq27a} is truncated at $k_{max}=335$, corresponding to the radius of 20 lattice constants; it was verified that such a selection leads to fully convergent calculations and further increase in $k_{max}$ does not influence the results.

For the purpose of numerical calculations, the dimensionless, reduced quantities are used and the energy scale is set by the quantity $k_{\rm B}T_{\rm D}^0$ (where $T^{0}_{\rm D}$ is the Debye temperature taken at $T=0$, $p=0$ and $h=0$). As an example, let us note the dimensionless pressure $\left(V_0/N\right) \left(p/\left(k_{\rm B}T_{\rm D}^0\right) \right)$. The Morse potential parameters $\alpha$ and $D$ were varied throughout the calculations, while the constant parameter $q=1$, occurring in Eq. \ref{eq10},  was accepted. The exponent in the power law for the NN exchange integral was chosen as $n=6$.

Let us commence the discussion of the results from the magnetic characteristics of the studied system.

One of the most crucial characteristics of the ferromagnet is its Curie temperature, which, in our approach, can be determined from the Eq.~\ref{eq26}. We make an assumption that the reference value of the NN exchange integral is the value at Curie temperature and zero external pressure, i.e. $J=J_1\left(T=T_{\rm C},p=0\right)$. Such a normalization is justified in the following manner: the usual way of determination of exchange integral involves the measurement of the Curie temperature and further application of the relation between this quantity and exchange integral (often a MFA formula is utilized, given in our paper by Eq.~\ref{eq36}, which is a linear dependence). In such a procedure the exchange integral is naturally determined at Curie temperature with $p$ neglected, so that we decided to use it as a reference value.  The value of the Curie temperature can be then conveniently normalized to the characteristic temperature $T^{0}_{\rm D}$. 

\begin{figure}[h!]
  \begin{center}
\includegraphics[scale=0.35]{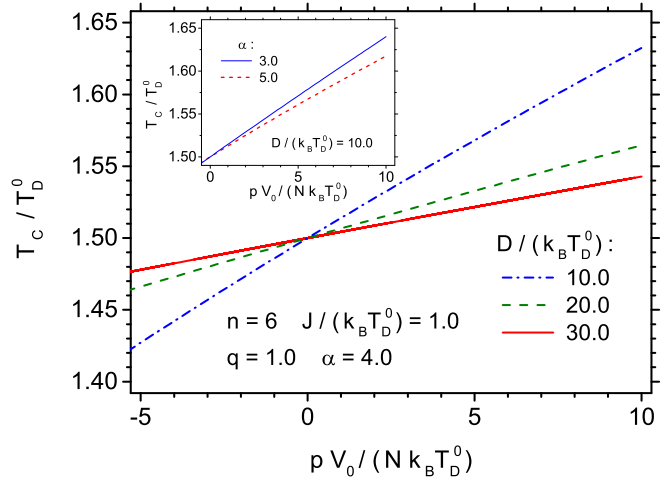}
  \end{center}
   \caption{\label{fig:fig1}Reduced Curie temperature of the system $T_{C}/T_{\rm D}^{0}$ as a function of the reduced pressure $pV_0/\left(Nk_{\rm B}T^{0}_{\rm D}\right)$, for various values of normalized parameter $D/\left(k_{\rm B}T^{0}_{\rm D}\right)$ describing the depth of the Morse potential (main plot) and for various values of the asymmetry parameter $\alpha$ for Morse potential (inset).}
\end{figure}

In Fig.~\ref{fig:fig1} we present the dependence of the normalized Curie temperature on the external pressure, plotted for the normalized exchange NN integral $J/\left(k_{\rm B}T^{0}_{\rm D}\right)=1.0$. The dependence in main panel is plotted for three representative values of the parameter $D$, which describes the depth of the Morse potential (see Eq.~\ref{eq3}). For the external pressure equal to 0, the normalized Curie temperature value of 1.5 is reached, regardless of the Morse potential parameters, which is the classical MFA result. This confirms the fact that the reference value of the exchange integral is the value reached at zero pressure and at the Curie temperature. 

It can be generally noticed that the Curie temperature is an increasing function of the external pressure. Such dependence is the sign of coupling between magnetic and lattice subsystems (as the exchange energy decreases with increasing interatomic distance). The form of the dependence is sensitive to the parameter $D$, as the deeper Morse potential reduces the influence of the pressure on the Curie temperature. On the other hand, relatively shallow lattice potential increases the sensitivity of $T_{\rm C}$ to pressure, making the dependence weakly non-linear, while this non-linearity vanishes for larger $D$. The inset in Fig.~\ref{fig:fig1} presents the pressure dependence of critical temperature for two values of parameter $\alpha$ describing the asymmetry of the Morse potential (see Eq.~\ref{eq3}) for relatively shallow potential well characterized by $D/\left(k_{\rm B}T_{\rm D}^{0}\right)=10.0$. It can be observed that for more asymmetric potential the sensitivity of the Curie temperature to pressure changes is reduced.

\begin{figure}[h!]
  \begin{center}
  \includegraphics[scale=0.35]{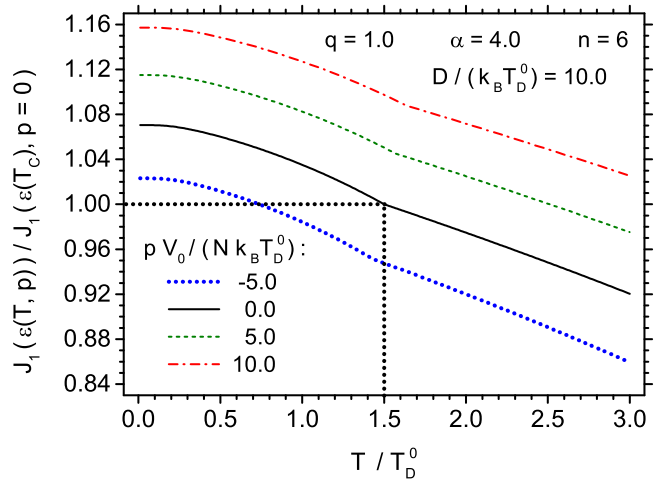}
  \end{center}
   \caption{\label{fig:fig2}Exchange integral between nearest neighbours normalized to its value at pressure $p=0$ and temperature equal to the critical temperature, as a function of the reduced temperature $T/T^{0}_{\rm D}$. The calculations are performed for various reduced pressures $pV_0/\left(Nk_{\rm B}T^{0}_{\rm D}\right)$. The kinks correspond to Curie temperatures.}
\end{figure}

The importance of indicating the conditions at which the reference exchange integral is determined can be justified on the basis of Fig.~\ref{fig:fig2}, which presents the dependence of the normalized NN exchange integral on the temperature for various external pressures. For the purpose of illustration, the relatively shallow Morse potential with $D/\left(k_{\rm B}T_{\rm D}^{0}\right)=10.0$ was selected. The normalized value of $J$ is equal to 1 at zero pressure and Curie temperature $1.5 T_{\rm D}^{0}$. The noticeable variability of $J$ as a function of pressure can be seen, with a characteristic kink at critical temperature. The dependence of $J$ on the temperature can be related to the temperature dependence of the relative deformation $\epsilon$ via Eq.~\ref{eq31}, which dependence will be shown and discussed in Fig.~\ref{fig:fig8}. The exchange integral is a decreasing function of interatomic distance, so that thermal expansion reduces the value of $J$.

\begin{figure}[h!]
  \begin{center}
   \includegraphics[scale=0.35]{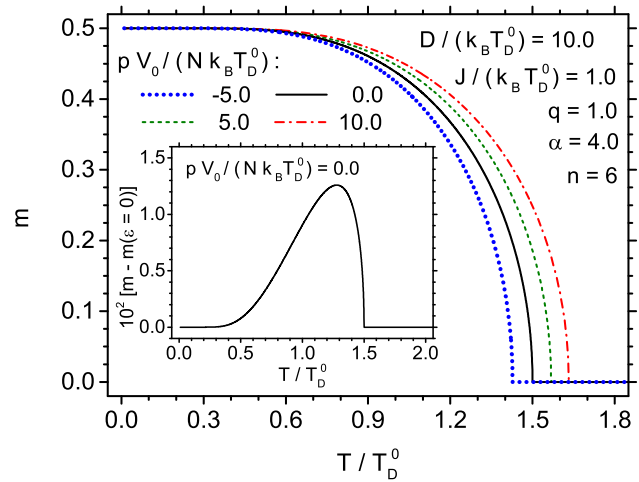}  
  \end{center}
   \caption{\label{fig:fig3}Magnetization as a function of the reduced temperature $T/T^{0}_{\rm D}$, for various reduced pressures $pV_0/\left(Nk_{\rm B}T^{0}_{\rm D}\right)$ (main plot). Difference in magnetizations calculated for zero pressure within present model and in the absence of magnetoelastic coupling as a function of the reduced temperature $T/T^{0}_{\rm D}$ (inset).}
\end{figure}

\begin{figure}[h!]
  \begin{center}
   \includegraphics[scale=0.35]{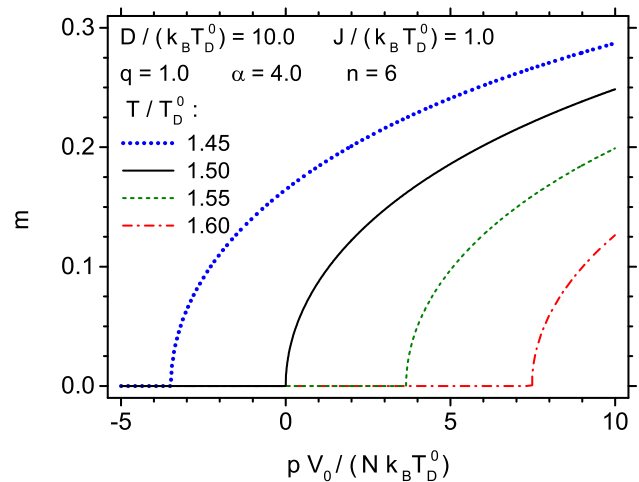} 
  \end{center}
   \caption{\label{fig:fig4}Magnetization as a function of the reduced pressure $pV_0/\left(Nk_{\rm B}T^{0}_{\rm D}\right)$, for various reduced temperatures $T/T^{0}_{\rm D}$ close to the Curie temperature of the system. }
\end{figure}

The variability of the Curie temperature under the influence of the external pressure can be followed also in Fig.~\ref{fig:fig3} presenting the temperature dependence of magnetization for various values of external pressure. The effect of shifting the Curie temperature by the pressure is clearly visible (see also Fig.~\ref{fig:fig1}). For various pressures, the dependences $m(T)$ remain monotonous and the change of magnetization due to the pressure changes is weakest close to the zero temperature (i.e. close to magnetic saturation). The effect of the pressure on the magnetization close to the Curie point will be separately shown in Fig.~\ref{fig:fig4}. 

It is also interesting to analyse the changes of the shape of $m(T)$ dependence under the influence of magnetoelastic interaction. The temperature dependence of the exchange integral $J_1$ implies also that the shape of the temperature dependence of magnetization, i.e. the function $m\left(T\right)$, is modified due to the coupling to the lattice. This is because in the second equation of state (Eq.~\ref{eq35}) we deal with the temperature-dependent values of $J_1$. These effects are traced in the inset to Fig.~\ref{fig:fig3}, which presents the difference between magnetization calculated for $p=0$ within the present model and the predictions of MFA with no magnetoelastic coupling. It is visible that the magnetization is slightly increased by the presence of the mentioned coupling and that the differences rise with the temperature and then drops close to the Curie temperature. For zero pressure the differences are rather limited (however, close to the Curie point, where the magnetization is small itself, the relative difference can become significant). It should be stressed that the differences are smallest for zero pressure. The dependence $m(T)$ without magnetoelastic coupling is completely insensitive to external pressure. Therefore, the differences between the curve plotted for $p=0$ and the lines depicting the functions for $p\neq 0$ seen in the main plot in Fig.~\ref{fig:fig3} are quite significant.

The temperature changes of magnetization are fastest in the vicinity of Curie temperature, for $T<T_{\rm C}$. The variation of the Curie temperature with pressure owing to coupling between magnetic and lattice system is therefore capable of causing the high sensitivity of magnetization to pressure at constant temperature close to the critical one. Such an effect is illustrated in Fig.~\ref{fig:fig4}, where pressure dependence of $m$ is plotted for several constant temperatures. Each of these temperatures corresponds to a Curie point for some pressure (compare with Fig.~\ref{fig:fig1} showing that Curie temperature is an increasing function of the pressure). At given temperature, for pressures lower than that required to reach a Curie point, the magnetization is equal to 0, since the system is in paramagnetic phase. The increase of pressure causes reaching the Curie point and a second-order, continuous transition to ferromagnetic ordering takes place. Further increase of pressure corresponds to the situation when the Curie temperatures are higher than the given temperature, so that the magnetization rises gradually. In this way, a continuous phase transition under isothermal conditions can take place, with pressure being a control parameter.

\begin{figure}[h!]
  \begin{center}
   \includegraphics[scale=0.35]{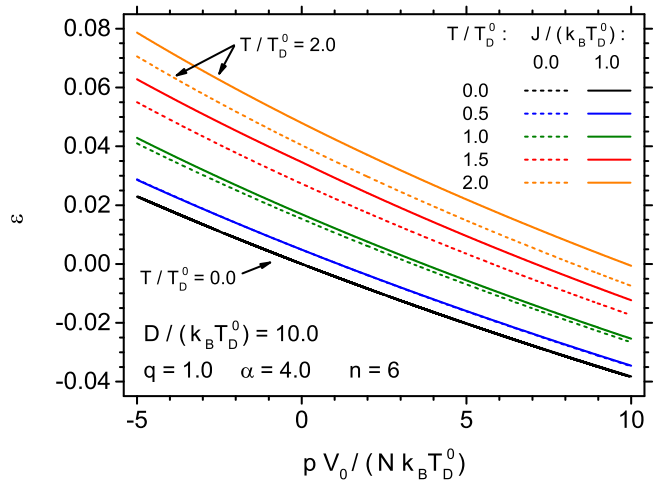} 
  \end{center}
   \caption{\label{fig:fig5}Isotropic volume deformation $\varepsilon$ as a function of the reduced pressure $pV_0/\left(Nk_{\rm B}T^{0}_{\rm D}\right)$ for various reduced temperatures $T/T^{0}_{\rm D}$. Dashed lines show the calculations performed in the absence of magnetic interactions, for $J/\left(k_{\rm B}T^{0}_{\rm D}\right)=0.0$; solid lines show the calculations performed in the presence of magnetic interactions, for $J/\left(k_{\rm B}T^{0}_{\rm D}\right)=1.0$. }
\end{figure}

\begin{figure}[h!]
  \begin{center}
   \includegraphics[scale=0.35]{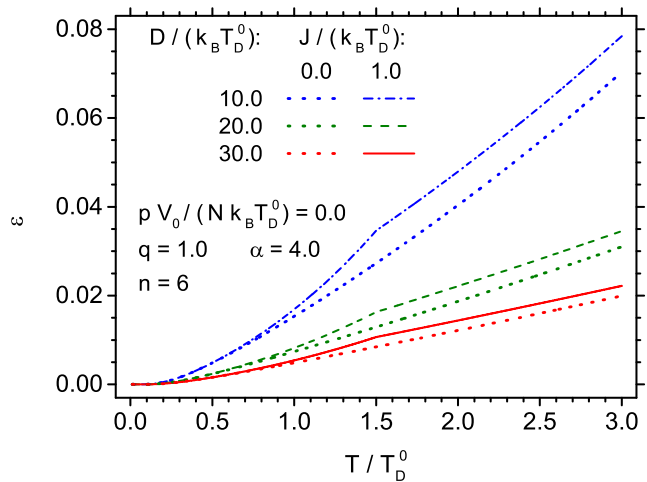} 
  \end{center}
   \caption{\label{fig:fig6}Isotropic volume deformation $\varepsilon$ as a function of reduced temperature $T/T^{0}_{\rm D}$. Dotted lines show the calculations performed in the absence of magnetic interactions, for $J/\left(k_{\rm B}T^{0}_{\rm D}\right)=0.0$; other lines show the calculations performed in the presence of magnetic interactions, for $J/\left(k_{\rm B}T^{0}_{\rm D}\right)=1.0$. Various values of normalized parameter $D/\left(k_{\rm B}T^{0}_{\rm D}\right)$ describing the depth of the Morse potential were accepted. The kinks correspond to Curie temperatures.  }
\end{figure}

\begin{figure}[h!]
  \begin{center}
   \includegraphics[scale=0.35]{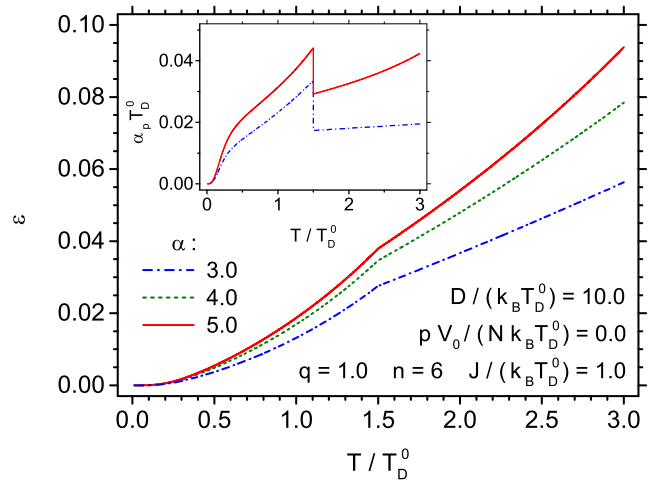} 
  \end{center}
   \caption{\label{fig:fig7}Isotropic volume deformation $\varepsilon$ as a function of reduced temperature $T/T^{0}_{\rm D}$ for various values of parameter $\alpha$ describing the asymmetry  of the Morse potential (main plot). Reduced thermal expansion coefficient $\alpha_{p}T^{0}_{\rm D}$  as a function of reduced temperature $T/T^{0}_{\rm D}$, for various values of parameter $\alpha$  (inset). The kinks and discontinuous jumps correspond to the Curie temperatures.}
\end{figure}

\begin{figure}[h!]
  \begin{center}
   \includegraphics[scale=0.35]{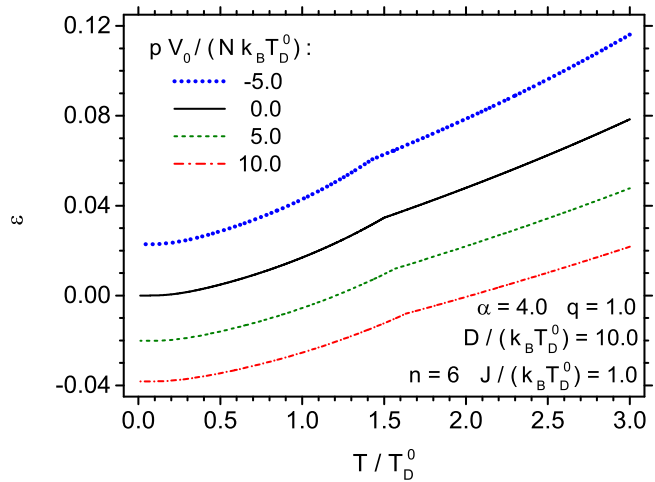} 
  \end{center}
   \caption{\label{fig:fig8}Isotropic volume deformation $\varepsilon$ as a function of the reduced temperature $T/T^{0}_{D}$ for various reduced external pressures $pV_0/\left(Nk_{\rm B}T^{0}_{\rm D}\right)$, in the presence of magnetic interactions with $J/\left(k_{\rm B}T^{0}_{\rm D}\right)=1.0$. The kinks correspond to the Curie temperature.}
\end{figure}

\begin{figure}[h!]
  \begin{center}
   \includegraphics[scale=0.35]{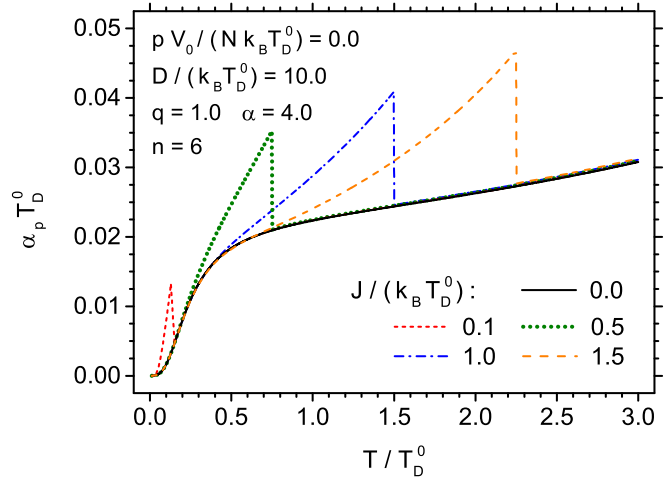} 
  \end{center}
   \caption{\label{fig:fig9}Reduced thermal expansion coefficient as a function of reduced temperature $T/T^{0}_{\rm D}$, for various exchange integrals $J/\left(k_{\rm B}T^{0}_{\rm D}\right)$, at pressure equal to 0. The discontinuous jumps occur at the Curie temperature. The solid line shows the results obtained in the absence of magnetic interactions.}
\end{figure}

\begin{figure}[h!]
  \begin{center}
   \includegraphics[scale=0.35]{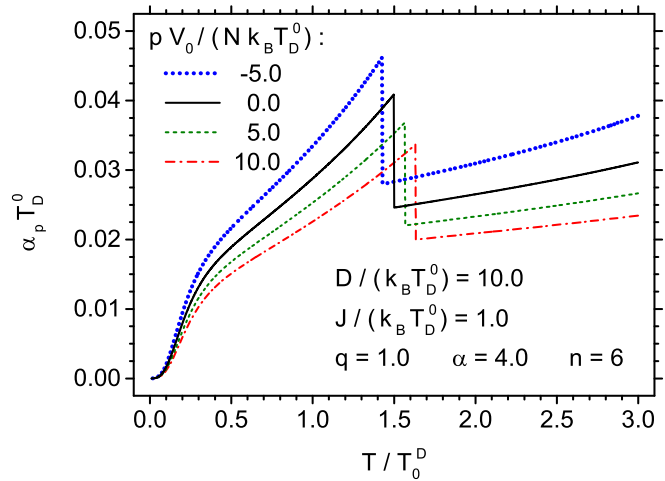} 
  \end{center}
   \caption{\label{fig:fig10}Reduced thermal expansion coefficient as a function of reduced temperature $T/T^{0}_{\rm D}$, for various reduced pressures $pV_0/\left(Nk_{\rm B}T^{0}_{\rm D}\right)$. The discontinuous jumps occurs at the Curie temperature.}
\end{figure}

\begin{figure}[h!]
  \begin{center}
   \includegraphics[scale=0.35]{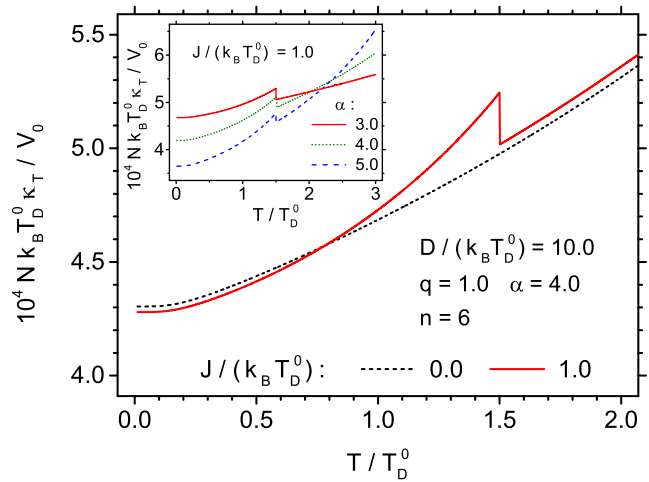} 
  \end{center}
   \caption{\label{fig:fig11}Reduced isothermal compressibility as a function of reduced temperature $T/T^{0}_{\rm D}$. Dashed line shows the calculations performed in the absence of magnetic interactions, for $J/\left(k_{\rm B}T^{0}_{\rm D}\right)=0.0$; solid line shows the calculations performed in the presence of magnetic interactions, for $J/\left(k_{\rm B}T^{0}_{\rm D}\right)=1.0$ (main plot). Reduced isothermal compressibility as a function of reduced temperature $T/T^{0}_{\rm D}$ for various values of the asymmetry parameter $\alpha$ for Morse potential (inset). The discontinuous jumps occurs at the Curie temperature.}
\end{figure}

\begin{figure}[h!]
  \begin{center}
   \includegraphics[scale=0.35]{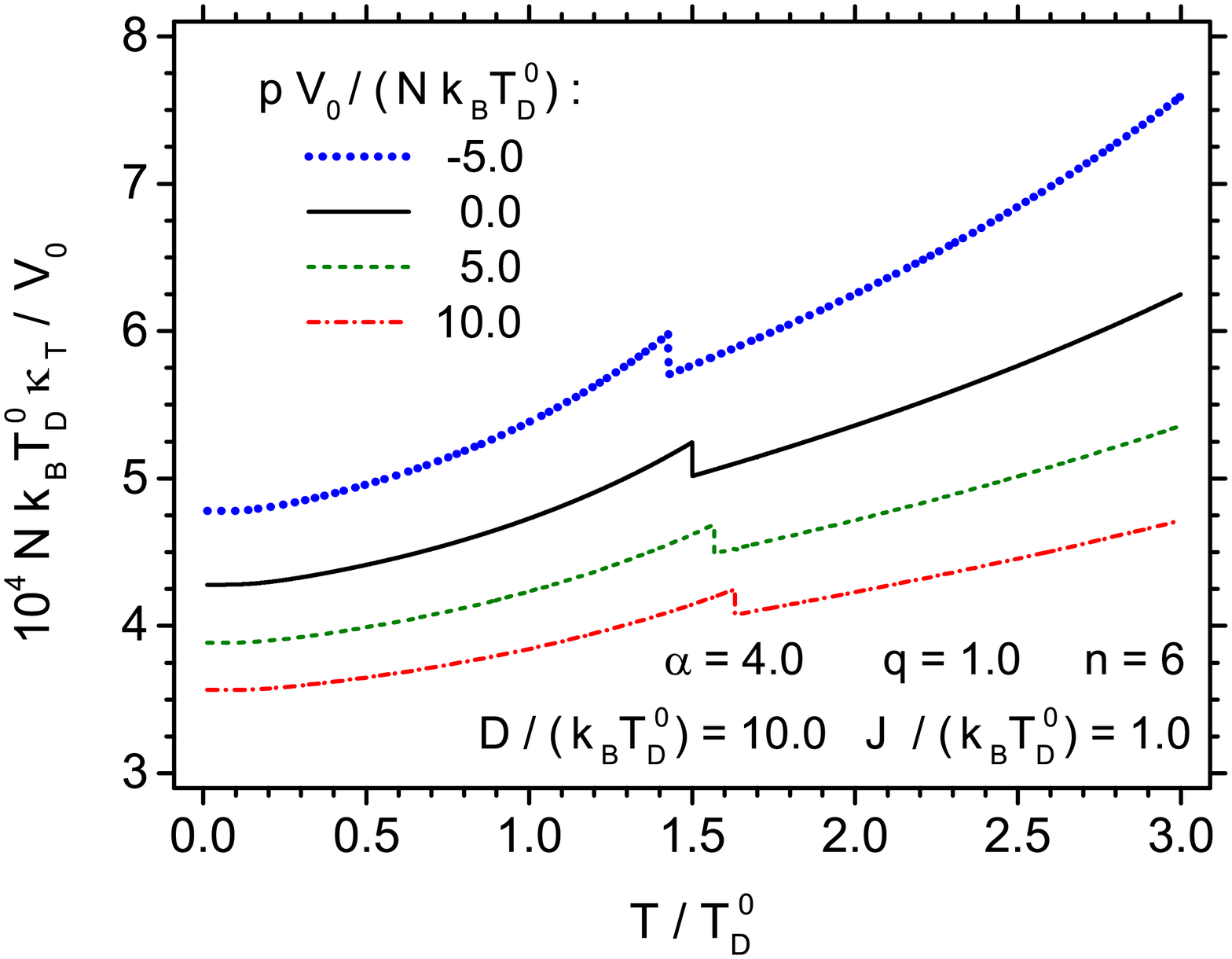} 
  \end{center}
   \caption{\label{fig:fig12}Reduced isothermal compressibility as a function of reduced temperature $T/T^{0}_{\rm D}$, for various reduced pressures $pV_0/\left(Nk_{\rm B}T^{0}_{\rm D}\right)$. The discontinuous jumps occur at the Curie temperature. }
\end{figure}

It is worth particular emphasis that not only the lattice subsystem influences the magnetic characteristics. Also the magnetic subsystem has a significant effect on non-magnetic properties.

Let us, therefore, discuss the lattice, mechanical properties of the studied system. The deformation of the system is described with the parameter $\epsilon$, which is temperature- and pressure-dependent. In Fig.~\ref{fig:fig5} the isotherms for various temperatures can be followed, i.e. the pressure dependence of the relative deformation $\epsilon$ at constant temperatures. The plot compares the results obtained in the absence of magnetism (dashed lines) and in the presence of magnetic subsystem with $J/\left(k_{\rm B}T^{0}_{\rm D}\right)=1.0$ (solid lines). It can be generally concluded that the low-temperature isotherms are insensitive to the presence of magnetism and magnetoelastic coupling (especially the one for zero temperature). The increase in temperature makes the $\epsilon(p)$ dependence more sensitive to the presence of the coupled magnetic subsystem, as the relative deformation in the same conditions is higher for $J>0$. The isotherms possess slightly non-linear, convex character. 

It is also highly interesting to follow the temperature dependence of relative deformation. Such data are shown in Fig.~\ref{fig:fig6}, where the dependence $\epsilon(T)$ is plotted for zero pressure. Various depths of the Morse potential are adopted and the data are collected to contrast the behaviour for the absence of magnetism (dotted lines) and in the presence of $J/\left(k_{\rm B}T^{0}_{\rm D}\right)=1.0$ (remaining lines). First, it is visible once more that the relative deformation in the presence of magnetism is larger than in the absence of it and the difference decreases when the Morse potential becomes deeper. At the lines plotted for $J/\left(k_{\rm B}T^{0}_{\rm D}\right)=1.0$ a kink can be observed close to $T/T^{0}_{\rm D}=1.5$. The position of the kink corresponds to the Curie temperature (see Fig.~\ref{fig:fig1}) of the system and separates the low-temperature ferromagnetic phase and high-temperature nonmagnetic phase. It can be seen that for $T>T_{\rm C}$ the shapes of the dependencies for $J=0$ and for $J>0$ are the same and the lines seem shifted in vertical direction. The larger depth of the Morse potential (parametrized by $D$) implies the smaller increase in relative deformation under the influence of the temperature, which effect can also be followed in Fig.~\ref{fig:fig6}. 

Another microscopic parameter which influences the lattice-related properties of a solid is the parameter $\alpha$ capturing the asymmetry of the Morse potential. Its effect on the temperature dependence of relative deformation is shown in Fig.~\ref{fig:fig7} (main plot). It can be stated that more symmetric potential (with lower $\alpha$) reduces the volume changes under the influence of the temperature. In the inset the reduced thermal expansion coefficient at constant pressure, $\alpha_{p}=\frac{1}{V}\left(\frac{\partial V}{\partial T}\right)_{p}$, is shown in the dimensionless form of $\alpha_{p}T^{0}_{\rm D}$. 

The influence of pressure on the relative deformation is presented in Fig.~\ref{fig:fig8}, where the temperature dependences of $\epsilon$ are plotted for various external pressures, both positive and negative ones. Let us note that for $p=0$ the relative deformation is equal to $\varepsilon_{\rm C}$  at $T=0$. It can be observed that the external pressure tends to shift the whole dependence $\epsilon(T)$ almost vertically. Some changes in the position of the kink occurring at the Curie temperature can be seen, corresponding to the pressure dependence of Curie temperature (see Fig.~\ref{fig:fig1}).

A crucial response function defining the properties of the solid is thermal expansion coefficient. This response function is defined as $\alpha_{p}=\frac{1}{V}\left(\frac{\partial V}{\partial T}\right)_{p}$, which can be also conveniently written as $\alpha_{p}=\frac{1}{1+\epsilon}\left(\frac{\partial \epsilon}{\partial T}\right)_{p}$. For the purpose of our further studies, the dimensionless quantity $\alpha_{p}T^{0}_{\rm D}$ can be introduced. As a derivative of deformation, the thermal expansion coefficient is much more sensitive to the influence of magnetic subsystem than $\epsilon(T)$ dependence itself. 

First, the dependence of the dimensionless thermal expansion coefficient on the temperature can be followed. In Fig.~\ref{fig:fig9} this quantity is plotted for various exchange integrals characterizing interspin interactions. In particular, the case of $J=0$ corresponds to lack of magnetic properties. In such case the coefficient $\alpha_{p}$ reaches the zero value at zero temperature and then monotonously increases.  In the whole range of temperatures $\alpha_{p}$ behaves continuously. The situation changes when the interaction with the magnetic subsystem is introduced by setting $J>0$. It is clearly visible that this interaction results in much faster increase of $\alpha_{p}$ in the low-temperature range, with a maximum value reached at the Curie point. At this point a discontinuity of thermal expansion coefficient occurs and the values for temperatures $T>T_{\rm C}$ follow the behaviour observed for $J=0$. It must be emphasized that the values of $\alpha_{p}$ in the vicinity of the Curie point (for $T<T_{\rm C}$) exceed the appropriate values for $J=0$ even by a factor of two. This proves the profound effect of coupling between magnetic and lattice system on thermal expansion. 

The example of the importance of the Morse potential parameters on behaviour of $\alpha_{p}$ can be followed in the inset in Fig.~\ref{fig:fig7}. The less symmetric Morse potential results in higher values of the thermal expansion coefficient themselves; moreover, the changes of $\alpha_{p}$ in the vicinity of the Curie point are more pronounced vs. temperature. 

It is also interesting to study the effect of the external pressure on the temperature dependence of thermal expansion coefficient, what is possible on the grounds of Fig.~\ref{fig:fig10}. It is evident the the positive (compressive) pressure reduces the value of $\alpha_{p}$ both for temperatures below and above Curie point. Also it tends to reduce the discontinuous jump of $\alpha_{p}$ at Curie temperature. It should be noted that the external pressure also shifts the Curie temperature itself (compare with Fig.~\ref{fig:fig1}), what is seen in Fig.~\ref{fig:fig10} as a shift in the discontinuity point. By comparison of Fig.~\ref{fig:fig10} and Fig.~\ref{fig:fig8}, once more it can be stated that the derivative quantity $\alpha_{p}$ is much more sensitive to the detailed changes than the function $\epsilon(T)$ itself.  

Another important quantity characterizing the volume response to changes of external pressure is the isothermal compressibility, which is defined as $\kappa_{T}=-\frac{1}{V}\left(\frac{\partial V}{\partial p}\right)_{T}$ and can be conveniently written in the form: $\kappa_{T}=-\frac{1}{1+\epsilon}\left(\frac{\partial \epsilon}{\partial p}\right)_{T}$. In our considerations the dimensionless quantity is $N k_{\rm B}T^{0}_{\rm D} \kappa_{T}/V_{0}$.

In Fig.~\ref{fig:fig11} (main plot) we compare the dependence of $\kappa{_T}$ on the temperature in the absence and in the presence of coupling between lattice and magnetic system for $p=0$. It is visible that for $J=0$ the compressibility is an increasing function of the temperature, with the initial slope at $T=0$ equal to 0. Comparison with the case of $J>0$ evidences that for very low temperatures the value of $\kappa_{T}$ is slightly decreased. However, the compressibility increases faster than for $J=0$ and, in the vicinity of Curie temperature, for $T<T_{\rm C}$, is significantly elevated in comparison to the case without magnetoelastic coupling (similar behaviour of thermal expansion coefficient was discussed above). The quantity $\kappa_{T}$ shows a discontinuous jump at the Cure point (similar to $\alpha_{p}$) and for $T>T_{\rm C}$ its temperature dependence resembles that observed for $J=0$. However, in the range of $T>T_{\rm C}$, values of compressibility for $J>0$ are still slightly higher than for $J=0$. 

The inset in Fig.~\ref{fig:fig11} presents the dependence of $\kappa_{T}$ on the temperature for various asymmetry parameters $\alpha$ of the Morse potential. It is evident that the limiting, low-temperature range of compressibility, is strongly $\alpha$-dependent whereas increasing asymmetry decreases $\kappa_{T}$. On the other hand, for higher $\alpha$, the compressibility rises faster with the temperature. Also the height of the jump at $T_{\rm C}$ is influenced by $\alpha$ - it is reduced by increasing asymmetry.

The effect of the external pressure on the temperature dependence of the compressibility can be followed in Fig.~\ref{fig:fig12}. The compressive pressure ($p>0$) decreases significantly the value of $\kappa_{T}$ and also influences the shape of $\kappa_{T}(T)$. Namely, the dependence becomes less convex when the pressure increases. The stretching pressure $p<0$ has an opposite effect. The discontinuous jump of $\kappa_{T}$ at $T_{\rm C}$ is visible, but its height is reduced by increasing external pressure. Once more, a shift of the Curie point with the pressure is clearly observed.

\section{Final remarks}
\label{sec4}

In the paper a fully self-consistent thermodynamic description of a ferromagnetic solid is presented, taking into consideration the presence of magnetoelastic coupling. The description is based on derivation of the total Gibbs free energy, dependent on the temperature $T$ and on both mechanical variables ($p$ and $V$) and magnetic ones ($h$ and $m$). Minimization of the Gibbs energy with respect to magnetization and elastic deformation of the solid leads to a pair of equations of state. Such equations allow the calculations of the mutual influence of lattice and magnetic properties. Moreover, the knowledge of the Gibbs energy sets the basis for studies of all interesting thermodynamic quantities. 

For the purpose of numerical calculations and illustration of the results, we considered a ferromagnetic solid with sc lattice and with the nearest-neighbour magnetic couplings following the power law as a function of interspin distance. For such system we performed extensive numerical calculations of both magnetic and lattice properties, revealing the importance of magnetoelastic coupling. For instance, the dependence of the Curie temperature and  magnetization on external pressure was studied. In turn, the influence of magnetic subsystem on thermal expansivity or compressibility was demonstrated. 

In our formalism we have assumed that the interatomic potential takes the form proposed by Morse \cite{Morse}; however, it can be generalized for arbitrary pair-wise interactions. The magnetic subsystem was characterized within Molecular Field Approximation for systems with the long-range interactions \cite{Szalowski}, yet a generalization involving more elaborate approximations is also possible. 

The method can be adopted for other isotropic bulk systems, including dilute alloys, disordered magnetics and/or higher spin models. In further extensions of the method the anisotropic volume deformation and anisotropic magnetic interactions should also be taken into account, with the prospects for the description of thin films.

\newpage
\appendix
\section{Exact calculation of the Debye integral}
\label{appendix1}

The integral appearing in Eqs.(\ref{eq12}) and (\ref{eq13}) can be presented in the form of:
\begin{equation}
\label{a1}
\int_{0}^{y_{\rm D}}{\frac{y^3}{e^{y}-1}\,dy}=\int_{0}^{\infty}{\frac{y^3}{e^{y}-1}\,d
y}-
\int_{y_{\rm D}}^{\infty}{\frac{y^3}{e^{y}-1}\,dy}=\frac{\pi^4}{15}-\int_{y_{\rm D}}^{\infty}{\frac{y^3}{e^{y}-1}\,dy}
\end{equation}
In Eq.(\ref{a1}) we deal with incomplete zeta function or "Debye function" which is given by the general expression \cite{Wood}:
\begin{equation}
\label{a2}
{\rm Z}_n\left(z\right)=\frac{1}{\left(n-1\right)!}\int_{z}^{\infty}{\frac{t^{n-1}}{e^{t}-1}\,dt}
\;\;\;\;\;\; \left(n=1,2,3,\dots\right)
\end{equation}
The "Debye function" can be expressed by the finite series of polylogarithms \cite{Wood}:
\begin{equation}
\label{a3}
{\rm Z}_n\left(z\right)=\sum_{k=0}^{n-1}{\rm Li}_{n-k}\left(e^{-z}\right) \frac{z^k}{k!}
\;\;\;\;\;\; \left(n=1,2,3,\dots\right)
\end{equation}
where 
\begin{equation}
\label{a4}
{\rm Li}_{s}\left(z\right)=\sum_{k=1}^{\infty}\frac{z^k}{k^s} 
\end{equation}
is the polylogarithm of order $s$ and argument $z$, extended by the process of analytic continuation.\\
The polylogarithm ${\rm Li}_{s}\left(z\right)$ for positive integer $s$ may be also expressed as a finite sum similar to Eq.(\ref{a3}) \cite{Wood}:
\begin{equation}
\label{a5}
{\rm Li}_{s}\left(e^{\mu}\right)=\sum_{k=0}^{s-1}{\rm Z}_{s-k}\left(- \mu \right)\frac{\mu^k}{k!} 
\;\;\;\;\;\; \left(s=1,2,3,\dots\right)
\end{equation}
Thus, the integral in Eq.(\ref{a1}) can be presented as:
\begin{equation}
\label{a6}
\int_{0}^{y_{\rm D}}{\frac{y^3}{e^{y}-1}\,dy}=\frac{\pi^4}{15}-3!\,{\rm Z}_4\left(y_{\rm D}\right)
\end{equation}
and, with the help of the expansion (\ref{a3}) takes the final form:
\begin{eqnarray}
\label{a7}
\int_{0}^{y_{\rm D}}{\frac{y^3}{e^{y}-1}\,dy}&=&
\frac{\pi^4}{15}-3!\sum_{k=0}^3{\rm Li}_{4-k}\left(e^{- y_{\rm D}}\right)
\frac{y_{\rm D}^k}{k!}\nonumber\\&=&
\frac{1}{15}\pi^4+y_{\rm D}^3 \ln \left(1- e^{-y_{\rm D}}\right)\nonumber\\
&-&3y_{\rm D}^2{\rm Li}_{2}\left(e^{-y_{\rm D}}\right) - 6y_{\rm D}{\rm Li}_{3}\left(e^{-y_{\rm D}}\right)-6{\rm Li}_{4}\left(e^{-y_{\rm D}}\right)
\end{eqnarray}
where ${\rm Li}_{1}\left(e^{-y_{\rm D}}\right)=-\ln \left(1-e^{-y_{\rm D}}\right)$.

\section{The case of nearest-neighbour magnetic interactions}
\label{appendix2}

A particularly interesting case is the case when the magnetic interactions are limited to nearest neighbours. Moreover, it can be conveniently assumed that their dependence on the distance between nearest-neighbour magnetic moments follows a power law:

\begin{equation}
\label{eq30}
J_1=J\left(\frac{r_1}{r_{1,C}}\right)^{-n},
\end{equation}
where $J$ is the exchange integral for NN at the Curie temperature $T_{\rm C}$, $h=0$ and $p=0$, whereas $r_{1,\rm C}$ is the NN distance in the same conditions.\\

By inserting Eqs.(\ref{eq15}) into (\ref{eq30}) we finally obtain $J_1$ as a function of $\varepsilon$:  
\begin{equation}
\label{eq31}
J_1=J\left(\frac{1+\varepsilon}{1+\varepsilon_{\rm C}}\right)^{-n/3}.
\end{equation}
Moreover, we have:
\begin{equation}
\label{eq32}
\left(\frac{\partial J_1}{\partial \varepsilon}\right)_{T}=-\frac{n}{3}J_1\frac{1}{1+\varepsilon}.
\end{equation}
Hence, the magnetic pressure (Eq. \ref{eq21}) for the considered case takes the value of:
\begin{equation}
\label{eq33}
p_{m}=-\frac{n}{6}\frac{N}{V_0}m^2\frac{1}{1+\varepsilon}J_1 z_{1}.
\end{equation}
In particular, for $T =0$, $p=0$ and $h=0$, the formula for magnetic pressure (Eq.~\ref{eq28}) takes the form of:
\begin{equation}
\label{eq34}
p_{m}=-\frac{n}{6}\frac{N}{V_0}S^2 J \left(1+\varepsilon_{\rm C}\right)^{n/3}z_1,
\end{equation}
and the equation \ref{eq29} yields:
\begin{eqnarray}
\label{eq34b}
&&\frac{1}{3}\frac{D}{k_{\rm B}T_{\rm D}^0}\alpha \frac{r_{1,0}}{r_{0}}
\sum_{k}^{}{z_k \frac{r_{k,0}}{r_{1,0}}\left[1-e^{-\alpha \left(\frac{r_{1,0}}{r_0}\frac{r_{k,0}}{r_{1,0}}-1\right)}\right]
e^{-\alpha \left(\frac{r_{1,0}}{r_0}\frac{r_{k,0}}{r_{1,0}}-1\right)}}\nonumber\\&&
+ 
\frac{n}{6}S^2 
\frac{J}{k_{\rm B}T_{\rm D}^0} \left(1+\varepsilon_{\rm C}\right)^{n/3}  z_1
=\frac{9}{8}\gamma_{\rm D}^0.
\end{eqnarray}

At the same time Eq.~\ref{eq22} is reduced to:
\begin{equation}
\label{eq35}
m=SB_S\left[S\beta \left(m J_1 z_1+h\right) \right],
\end{equation}
while the Curie temperature (Eq.~\ref{eq26}) amounts to:
\begin{equation}
\label{eq36}
k_{\rm B}T_{\rm C}=\frac{S\left(S+1\right)}{3} J_1 z_1.
\end{equation}


\begin{thebibliography}{10}
\expandafter\ifx\csname url\endcsname\relax
  \def\url#1{\texttt{#1}}\fi
\expandafter\ifx\csname urlprefix\endcsname\relax\def\urlprefix{URL }\fi
\expandafter\ifx\csname href\endcsname\relax
  \def\href#1#2{#2} \def\path#1{#1}\fi

\bibitem{Stanley}
H.~E. Stanley, Introduction to Phase Transitions and Critical Phenomena, Oxford
  University Press, 1971.

\bibitem{Balcerzak1998}
T.~Balcerzak, A rigorous equation of state for magnetic systems, Journal of
  Magnetism and Magnetic Materials 177 (1998) 771--772.
\newblock \href {http://dx.doi.org/10.1016/S0304-8853(97)00291-6}
  {\path{doi:10.1016/S0304-8853(97)00291-6}}.

\bibitem{Salinas1973}
S.~R. Salinas, On the one-dimensional compressible {I}sing model, Journal of
  Physics A: Mathematical, Nuclear and General 6~(10) (1973) 1527.
\newblock \href {http://dx.doi.org/10.1088/0305-4470/6/10/011}
  {\path{doi:10.1088/0305-4470/6/10/011}}.

\bibitem{Bergman1976}
D.~J. Bergman, B.~I. Halperin, Critical behavior of an {I}sing model on a cubic
  compressible lattice, Phys. Rev. B 13 (1976) 2145--2175.
\newblock \href {http://dx.doi.org/10.1103/PhysRevB.13.2145}
  {\path{doi:10.1103/PhysRevB.13.2145}}.

\bibitem{Chakrabarti1977}
B.~Chakrabarti, Critical behaviour of a compressible {I}sing model, Solid State
  Communications 23~(9) (1977) 683 -- 685.
\newblock \href {http://dx.doi.org/10.1016/0038-1098(77)90549-X}
  {\path{doi:10.1016/0038-1098(77)90549-X}}.

\bibitem{Chakrabarti1979}
B.~K. Chakrabarti, B.~P. Barua, S.~K. Sinha, Critical temperature of a
  compressible {I}sing magnet, physica status solidi (b) 94~(2) (1979)
  621--625.
\newblock \href {http://dx.doi.org/10.1002/pssb.2220940234}
  {\path{doi:10.1002/pssb.2220940234}}.

\bibitem{Henriques1987}
V.~B. Henriques, S.~R. Salinas, Effective spin hamiltonians for compressible
  {I}sing models, Journal of Physics C: Solid State Physics 20~(16) (1987)
  2415.
\newblock \href {http://dx.doi.org/10.1088/0022-3719/20/16/014}
  {\path{doi:10.1088/0022-3719/20/16/014}}.

\bibitem{Diep1999}
E.~H. Boubcheur, H.~T. Diep, Effect of elastic interaction on critical behavior
  of three-dimensional {I}sing model, Journal of Applied Physics 85~(8) (1999)
  6085--6087.
\newblock \href {http://dx.doi.org/10.1063/1.369090}
  {\path{doi:10.1063/1.369090}}.

\bibitem{Diep2000}
P.~Massimino, H.~T. Diep, Effect of magnetoelastic interactions on the phase
  transition of two-dimensional {I}sing spin system, Journal of Applied Physics
  87~(9) (2000) 7043--7045.
\newblock \href {http://dx.doi.org/10.1063/1.372925}
  {\path{doi:10.1063/1.372925}}.

\bibitem{Boubcheur2001}
E.~Boubcheur, P.~Massimino, H.~Diep, Effects of magnetoelastic coupling:
  critical behavior and structure deformation, Journal of Magnetism and
  Magnetic Materials 223~(2) (2001) 163 -- 168.
\newblock \href {http://dx.doi.org/10.1016/S0304-8853(00)00752-6}
  {\path{doi:10.1016/S0304-8853(00)00752-6}}.

\bibitem{Landau2005}
L.~Cannavacciuolo, D.~P. Landau, Critical behavior of the three-dimensional
  compressible {I}sing antiferromagnet at constant volume: {A} {M}onte {C}arlo
  study, Phys. Rev. B 71 (2005) 134104.
\newblock \href {http://dx.doi.org/10.1103/PhysRevB.71.134104}
  {\path{doi:10.1103/PhysRevB.71.134104}}.

\bibitem{Li2010}
P.~Li, Y.~Chen, Magnetoelastic instability in {I}sing-like models, Physics
  Letters A 374~(3) (2010) 453 -- 456.
\newblock \href {http://dx.doi.org/10.1016/j.physleta.2009.10.085}
  {\path{doi:10.1016/j.physleta.2009.10.085}}.

\bibitem{Chakrabarti1980}
B.~K. Chakrabarti, Critical behaviour of compressible dilute {I}sing systems,
  Journal of Physics C: Solid State Physics 13~(24) (1980) 4505.
\newblock \href {http://dx.doi.org/10.1088/0022-3719/13/24/013}
  {\path{doi:10.1088/0022-3719/13/24/013}}.

\bibitem{Chakrabarti1982}
B.~K. Chakrabarti, Critical behaviour of compressible dilute ising systems-a
  {M}onte {C}arlo study, Journal of Physics C: Solid State Physics 15~(33)
  (1982) L1195.
\newblock \href {http://dx.doi.org/10.1088/0022-3719/15/33/004}
  {\path{doi:10.1088/0022-3719/15/33/004}}.

\bibitem{Chakrabarti1980b}
B.~K. Chakrabarti, G.~A. Gehring, Critical behaviour of quantum compressible
  {I}sing models, Journal of Physics C: Solid State Physics 13~(24) (1980)
  4495.
\newblock \href {http://dx.doi.org/10.1088/0022-3719/13/24/012}
  {\path{doi:10.1088/0022-3719/13/24/012}}.

\bibitem{Pytte1965}
E.~Pytte, Spin-phonon interactions in a {H}eisenberg ferromagnet, Annals of
  Physics 32~(3) (1965) 377--403.
\newblock \href {http://dx.doi.org/10.1016/0003-4916(65)90139-9}
  {\path{doi:10.1016/0003-4916(65)90139-9}}.

\bibitem{Barma}
M.~Barma, Phonon-induced phase transition in a classical {H}eisenberg chain,
  Phys. Rev. B 12 (1975) 2710--2715.
\newblock \href {http://dx.doi.org/10.1103/PhysRevB.12.2710}
  {\path{doi:10.1103/PhysRevB.12.2710}}.

\bibitem{Diep2002}
V.~T. Ngo, H.~T. Diep, Monte {C}arlo study of surface-frustrated heisenberg
  thin films with magnetoelastic coupling: {A}n off-lattice model, Journal of
  Applied Physics 91~(10) (2002) 8399--8401.
\newblock \href {http://dx.doi.org/10.1063/1.1456443}
  {\path{doi:10.1063/1.1456443}}.

\bibitem{Liarte}
D.~B. Liarte, S.~R. Salinas, C.~S.~O. Yokoi, Compressible
  {S}herrington-{K}irkpatrick spin-glass model, Journal of Physics A:
  Mathematical and Theoretical 42~(20) (2009) 205002.
\newblock \href {http://dx.doi.org/10.1088/1751-8113/42/20/205002}
  {\path{doi:10.1088/1751-8113/42/20/205002}}.

\bibitem{Sobkowicz1996}
B.~Sobkowicz, Markand~Chakraborty, Ising model with frustration, elasticity,
  and competing interactions, Journal of Statistical Physics 83~(3) (1996)
  739--749.
\newblock \href {http://dx.doi.org/10.1007/BF02183746}
  {\path{doi:10.1007/BF02183746}}.

\bibitem{Vecchini}
C.~Vecchini, M.~Poienar, F.~Damay, O.~Adamopoulos, A.~Daoud-Aladine, A.~Lappas,
  J.~M. Perez-Mato, L.~C. Chapon, C.~Martin, Magnetoelastic coupling in the
  frustrated antiferromagnetic triangular lattice {CuMnO}$_2$, Phys. Rev. B 82
  (2010) 094404.
\newblock \href {http://dx.doi.org/10.1103/PhysRevB.82.094404}
  {\path{doi:10.1103/PhysRevB.82.094404}}.

\bibitem{Zorko}
A.~Zorko, J.~Kokalj, M.~Komelj, O.~Adamopoulos, H.~Luetkens, D.~Ar\v{c}on,
  A.~Lappas, Magnetic inhomogeneity on a triangular lattice: the
  magnetic-exchange versus the elastic energy and the role of disorder,
  Scientific Reports 5 (2015) 9272.
\newblock \href {http://dx.doi.org/10.1038/srep09272}
  {\path{doi:10.1038/srep09272}}.

\bibitem{Bergman1973}
D.~J. Bergman, Y.~Imry, L.~Gunther, Exactly soluble magnetoelastic lattice with
  a magnetic phase transition, Journal of Statistical Physics 7~(4) (1973)
  337--360.
\newblock \href {http://dx.doi.org/10.1007/BF01014909}
  {\path{doi:10.1007/BF01014909}}.

\bibitem{Strecka2012}
J.~Stre{\v{c}}ka, O.~Rojas, S.~M. de~Souza, Spontaneous distortion in the
  spin-1/2 {I}sing-{H}eisenberg model on decorated planar lattices with a
  magnetoelastic coupling, The European Physical Journal B 85~(2) (2012) 1--12.
\newblock \href {http://dx.doi.org/10.1140/epjb/e2011-20733-5}
  {\path{doi:10.1140/epjb/e2011-20733-5}}.

\bibitem{Strecka2012b}
J.~Stre{\v{c}}ka, O.~Rojas, S.~de~Souza, Spin-phonon coupling induced
  frustration in the exactly solved spin-1/2 {I}sing model on a decorated
  planar lattice, Physics Letters A 376~(3) (2012) 197--202.
\newblock \href
  {http://dx.doi.org/http://dx.doi.org/10.1016/j.physleta.2011.11.008}
  {\path{doi:http://dx.doi.org/10.1016/j.physleta.2011.11.008}}.

\bibitem{Wojtczak}
J.~Rutkowski, L.~Wojtczak, \v{S}. Zajac,
  \href{http://przyrbwn.icm.edu.pl/APP/PDF/118/a118z5p012.pdf}{Pressure
  influence on the curie temperature}, Acta Physica Polonica A 118~(5) (2010)
  745--746.
\newline\urlprefix\url{http://przyrbwn.icm.edu.pl/APP/PDF/118/a118z5p012.pdf}

\bibitem{Leger}
J.~M. Leger, C.~Loriers-Susse, B.~Vodar, Pressure effect on the curie
  temperatures of transition metals and alloys, Phys. Rev. B 6 (1972)
  4250--4261.
\newblock \href {http://dx.doi.org/10.1103/PhysRevB.6.4250}
  {\path{doi:10.1103/PhysRevB.6.4250}}.

\bibitem{Gehring}
G.~A. Gehring, Pressure-induced quantum phase transitions, EPL 82~(6) (2008)
  60004.
\newblock \href {http://dx.doi.org/10.1209/0295-5075/82/60004}
  {\path{doi:10.1209/0295-5075/82/60004}}.

\bibitem{Radomska2000}
A.~Radomska, T.~Balcerzak,
  \href{http://przyrbwn.icm.edu.pl/APP/PDF/98/a098z1p09.pdf}{Calculations of
  {EuTe} magnetic phase diagram for external pressure}, Acta Physica Polonica A
  98~(1-2) (2000) 83--91.
\newline\urlprefix\url{http://przyrbwn.icm.edu.pl/APP/PDF/98/a098z1p09.pdf}

\bibitem{Radomska2001}
A.~Radomska, T.~Balcerzak, The magnetic phase diagram of {EuTe} for high
  external pressure, physica status solidi (b) 225~(1) (2001) 229--236.
\newblock \href
  {http://dx.doi.org/10.1002/(SICI)1521-3951(200105)225:1<229::AID-PSSB229>3.0.CO;2-L}
  {\path{doi:10.1002/(SICI)1521-3951(200105)225:1<229::AID-PSSB229>3.0.CO;2-L}}.

\bibitem{Radomska2003}
A.~Radomska, T.~Balcerzak, Theoretical studies of model thin {EuTe} films with
  surface elastic stresses, Central European Journal of Physics 1~(1) (2003)
  100--117.
\newblock \href {http://dx.doi.org/10.2478/BF02475555}
  {\path{doi:10.2478/BF02475555}}.

\bibitem{Sollinger2010}
W.~S\"ollinger, W.~Heiss, R.~T. Lechner, K.~Rumpf, P.~Granitzer, H.~Krenn,
  G.~Springholz, Exchange interactions in europium monochalcogenide magnetic
  semiconductors and their dependence on hydrostatic strain, Phys. Rev. B 81
  (2010) 155213.
\newblock \href {http://dx.doi.org/10.1103/PhysRevB.81.155213}
  {\path{doi:10.1103/PhysRevB.81.155213}}.

\bibitem{Amaral2004}
V.~Amaral, J.~Amaral, Magnetoelastic coupling influence on the magnetocaloric
  effect in ferromagnetic materials, Journal of Magnetism and Magnetic
  Materials 272-276 (2004) 2104 -- 2105.
\newblock \href {http://dx.doi.org/10.1016/j.jmmm.2003.12.870}
  {\path{doi:10.1016/j.jmmm.2003.12.870}}.

\bibitem{Sigrist2005}
M.~Matsumoto, M.~Sigrist, Ehrenfest relations and magnetoelastic effects in
  field-induced ordered phases, Journal of the Physical Society of Japan 74~(8)
  (2005) 2310--2316.
\newblock \href {http://dx.doi.org/10.1143/JPSJ.74.2310}
  {\path{doi:10.1143/JPSJ.74.2310}}.

\bibitem{Alho2010}
B.~P. Alho, N.~A. de~Oliveira, V.~S.~R. de~Sousa, E.~J.~R. Plaza, A.~M.~G.
  Carvalho, P.~J. von Ranke, The influence of the magnetoelastic interaction on
  the magnetocaloric effect in ferrimagnetic systems: a theoretical
  investigation, Journal of Physics: Condensed Matter 22~(48) (2010) 486008.
\newblock \href {http://dx.doi.org/10.1088/0953-8984/22/48/486008}
  {\path{doi:10.1088/0953-8984/22/48/486008}}.

\bibitem{Singh2013}
N.~Singh, R.~Arr{\'{o}}yave, Magnetocaloric effects in {Ni-Mn-Ga-Fe} alloys
  using {M}onte {C}arlo simulations, Journal of Applied Physics 113~(18).
\newblock \href {http://dx.doi.org/10.1063/1.4803544}
  {\path{doi:10.1063/1.4803544}}.

\bibitem{Morse}
P.~M. Morse, Diatomic molecules according to the wave mechanics. {II}.
  {V}ibrational levels, Phys. Rev. 34 (1929) 57--64.
\newblock \href {http://dx.doi.org/10.1103/PhysRev.34.57}
  {\path{doi:10.1103/PhysRev.34.57}}.

\bibitem{Girifalco}
L.~A. Girifalco, V.~G. Weizer, Application of the morse potential function to
  cubic metals, Phys. Rev. 114 (1959) 687--690.
\newblock \href {http://dx.doi.org/10.1103/PhysRev.114.687}
  {\path{doi:10.1103/PhysRev.114.687}}.

\bibitem{Lincoln}
R.~C. Lincoln, K.~M. Koliwad, P.~B. Ghate, Morse-potential evaluation of
  second- and third-order elastic constants of some cubic metals, Phys. Rev.
  157 (1967) 463--466.
\newblock \href {http://dx.doi.org/10.1103/PhysRev.157.463}
  {\path{doi:10.1103/PhysRev.157.463}}.

\bibitem{Balcerzak}
T.~Balcerzak, K.~Sza\l{}owski, M.~Ja\v{s}\v{c}ur, A self-consistent
  thermodynamic model of metallic systems. {A}pplication for the description of
  gold, Journal of Applied Physics 116~(4).
\newblock \href {http://dx.doi.org/10.1063/1.4891251}
  {\path{doi:10.1063/1.4891251}}.

\bibitem{Matsui}
M.~Matsui, High temperature and high pressure equation of state of gold,
  Journal of Physics: Conference Series 215~(1) (2010) 012197.
\newblock \href {http://dx.doi.org/10.1088/1742-6596/215/1/012197}
  {\path{doi:10.1088/1742-6596/215/1/012197}}.

\bibitem{Gruneisen}
E.~Gr\"{u}neisen, Theorie des festen {Z}ustandes einatomiger {E}lemente,
  Annalen der Physik 344~(12) (1912) 257--306.
\newblock \href {http://dx.doi.org/10.1002/andp.19123441202}
  {\path{doi:10.1002/andp.19123441202}}.

\bibitem{Krivtsov2011}
V.~A. Krivtsov, A. M.and~Kuz'kin, Derivation of equations of state for ideal
  crystals of simple structure, Mechanics of Solids 46~(3) (2011) 387--399.
\newblock \href {http://dx.doi.org/10.3103/S002565441103006X}
  {\path{doi:10.3103/S002565441103006X}}.

\bibitem{Wallace}
D.~C. Wallace, Thermodynamics of Crystals, Wiley, 1972.

\bibitem{Dubinov2008}
A.~A. Dubinov, A. E.and~Dubinova, Exact integral-free expressions for the
  integral {D}ebye functions, Technical Physics Letters 34~(12) (2008)
  999--1001.
\newblock \href {http://dx.doi.org/10.1134/S106378500812002X}
  {\path{doi:10.1134/S106378500812002X}}.

\bibitem{Wood}
D.~Wood, \href{http://www.cs.kent.ac.uk/pubs/1992/110}{The Computation of
  Polylogarithms. {T}echnical report 15-92}, University of Kent Computing
  Laboratory, Canterbury, UK, 1992.
\newline\urlprefix\url{http://www.cs.kent.ac.uk/pubs/1992/110}

\bibitem{Szalowski}
K.~Sza\l{}owski, T.~Balcerzak, Phase diagrams of a model diluted fcc magnet
  with arbitrary spin and modified {RKKY} interaction: {I}nfluence of external
  magnetic field and structural short-range order, Phys. Rev. B 77 (2008)
  115204.
\newblock \href {http://dx.doi.org/10.1103/PhysRevB.77.115204}
  {\path{doi:10.1103/PhysRevB.77.115204}}.

\bibitem{Wiren}
Z.~Q. Wiren,
  \href{http://ir.library.oregonstate.edu/xmlui/bitstream/handle/1957/8752/ZQW_thesis.pdf}{Exchange
  interaction studies in magnetic semiconductors by neutron scattering}, Ph.D.
  thesis, Oregon State University (2008).
\newline\urlprefix\url{http://ir.library.oregonstate.edu/xmlui/bitstream/handle/1957/8752/ZQW_thesis.pdf}

\bibitem{Szuszkiewicz2006}
W.~Szuszkiewicz, E.~Dynowska, B.~Witkowska, B.~Hennion, Spin-wave measurements
  on hexagonal $\mathrm{MnTe}$ of $\mathrm{NiAs}$-type structure by inelastic
  neutron scattering, Phys. Rev. B 73 (2006) 104403.
\newblock \href {http://dx.doi.org/10.1103/PhysRevB.73.104403}
  {\path{doi:10.1103/PhysRevB.73.104403}}.

\end{thebibliography}

\end{document}